# Conductance spectra of metallic nanotube bundles

Wei Ren[1], C. T. Chan[1], T. H. Cho[2], T. C. Leung[2], Jian Wang[3], Hong Guo[4], and Ping Sheng[1]

[1]Department of Physics, Hong Kong University of Science and Technology

Clear Water Bay, Kowloon, Hong Kong, China

[2]Department of Physics, National Chung Cheng University

Chia-Yi, Taiwan, Republic of China

[3]Center of Theoretical and Computational Physics and Department of Physics

The University of Hong Kong, Pokfulam Road, Hong Kong, China

[4]Department of Physics, McGill University, Montreal, Quebec, Canada

**Abstract**

We report a first principles analysis of electronic transport characteristics for $(n,n)$ carbon nanotube bundles. When $n$ is not a multiple of 3, inter-tube coupling causes universal conductance suppression near Fermi level regardless of the rotational arrangement of individual tubes. However, when $n$ is a multiple of 3, the bundles exhibit a diversified conductance dependence on the orientation details of the constituent tubes. The total energy of the bundle is also sensitive to the orientation arrangement only when $n$ is a multiple of 3. All the transport properties and band structures can be well understood from the symmetry consideration of whether the rotational symmetry of the individual tubes is commensurate with that of the bundle.

Single walled carbon nantubes (SWNT) can form a bundle with a triangular cross-sectional lattice in a self-organized manner, a good example of hierarchical solids. The number of SWNTs in a close-packed bundle ranges from a few to hundreds. It has been known that the bundles containing parallel individual nanotubes possess superb mechanic[1], thermal[2] and electronic[3-5] properties. Since nanotube bundles are ubiquitous in nanotube synthesis reaction, they have attracted great attention both experimentally and theoretically. Very recently it was determined that the minimal bundling number is between 3 and 8 in vertically aligned SWNT films[6]. For these bundles containing small number of tubes, the 1D properties of SWNT can still be attainable. This is in contrast to bundles with large number of tubes where bulk properties might dominate.

SWNT bundles as nanowires have exciting prospect for nanoelectronics. Particularly metallic bundles have been proposed theoretically and experimentally[7] to outperform conventional copper wires for interconnect application. However, the electronic transport physics of metallic SWNT nanotube bundles are more complex than that of a single tube. It results from the multitudes of numerous possible rotations of tubes with respect to each other, as well as the concomitant symmetry breaking effects in presence of delicate coupling between the tubes. On the theoretical side, quantum transport properties of nanotube bundles are difficult to investigate from first principles because of the large number of atoms involved in a bundle, and it is the purpose of this work to fill this gap. In particular, we focus on metallic bundles made of armchair SWNTs and calculate their conductance spectra using a state-of-the-art *ab initio* technique. We find the transport properties of bundles made of ($n,n$) armchair SWNTs to be well classified into several categories depending on the chirality index *n* of the tubes in the bundle. Each category is characterized by whether no gap, a partial gap, or a complete gap exists at the Fermi level ($E_f$) of the conductance spectra. These gap structures are the direct consequences of geometric symmetry of the SWNT's inside the bundle, and they give substantial influence to the conductance of the bundle.

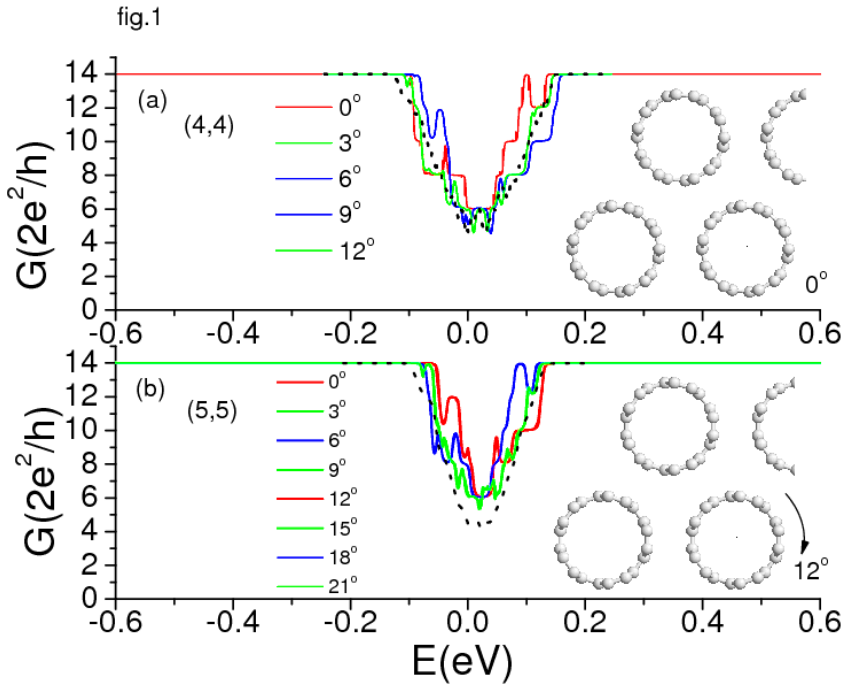

**FIG. 1. (Color online) Electronic conductance of 7-tube bundles consisting of (4,4) and (5,5) SWNTs, for different collective rotational angles (solid) and randomly orientated (dotted) bundles. Insets show the schematic end views of the SWNT bundles. Upper inset: the nanotubes are aligned along a horizontal axis in the cross-section plane; lower inset: all tubes are rotated by 12° clockwise around tube axis collectively, and categorized as an aligned bundle.**

A bundle has more electronic conduction channels than a single nanotube. Without considering inter-tube coupling, the conductance of the bundle should simply scale with the number of tubes in the bundle. When there is tube-tube interaction, the coupling can induce symmetry breaking that mixes the π and π* bands, leading to possible opening of band gaps near the Fermi level. Therefore the total conductance of a bundle may not be the summation of the conductance of all participating tubes. Instead the orientation becomes important because how adjacent tubes interact with each other dominates the electronic transport. Our work will systematically examine the characteristics of electronic structure and total energy of the bundles, and we correlate the transport properties to these characteristics in a general manner. The results can be explained qualitatively by symmetry considerations. Our quantum transport and band structure calculations of SWNT armchair bundles are accomplished by a first-principle real space electron transport package MATDCAL[8, 9]. In this technique, density functional theory (DFT) is carried out with the Keldysh nonequilibrium Green's function (NEGF) formalism. The details of the NEGF-DFT formalism can be found elsewhere[10, 11]. In this work, we adopt standard non-local, norm-conserving pseudopotentials[12] for the atomic cores and *s, p, d* single-zeta real-space basis set within the local density approximation. We calculated bundles with a finite cross section but having infinite length along the tube axis. For simplicity, individual tubes are assumed to have no impurities or defects so that all effects can be attributed to the coupling between the tubes. We assume a bundle contains seven nanotubes, with one tube in the middle surrounded by six other tubes forming a hexagonal pattern. This configuration mimics the close-packed structure of realistic nanotube bundles[6]. The nearest wall-to-wall distance of two adjacent tubes is taken to be 3.33 Å unless otherwise stated. Other kinds of bundles are also investigated (see below). We have studied bundles in which the tubes are aligned with various fixed rotational angles and also randomly-orientated ones. In the aligned bundle, all the tubes are orientated with a certain angle with respect to a pre-specified axis; in randomly-orientated bundle, each tube has a different orientation angle drawn from a random-number generator[13]. Finally, we have also employed the plane-wave basis DFT code VASP4.6.26[14] to calculate band structure and total energy of a *periodic array* of bundles. Energy cutoff of 286.6 eV, US potential and Ceperley and Alder (CA) local exchange-correlation functional were applied. The Brillouin zone was sampled with 6x6x12 k points of a Monkhorst-Pack grid. All the atoms in the unit cell and cell parameters were fully relaxed. The band structures obtained by VASP on the bulk are in good agreement with that obtained by MATDCAL on bundles with finite cross section.

We first consider the electric conductance of the (*4,4*) and (*5,5*) tube bundles. For a pristine single (*n,n*) armchair metallic nanotube, there are two Bloch states crossing $E_f$ of the tube. Therefore in the ballistic transport regime one obtains a conductance $G=2G_o$ for one tube where the conductance quanta $G_o=2e^2/h$. Since there are seven tubes in a bundle, we expect a total equilibrium conductance of $14G_o$. Fig.1 shows that away from the immediate neighborhood of $E_f$, the calculated conductance indeed has a value of $14G_o$. But within a 0.2 eV range near $E_f$, the conductance of the bundle is strongly suppressed by a factor up to 2/3. Such a large suppression of the bundle's conductance is caused by inter-tube coupling, which induces *pseudogaps* in the electronic structure near $E_f$ [3], with modified density of states (DOS) that reduces transmission coefficients of the conduction channels. Here we use the term "*pseudogap*" to indicate the situation shown in Fig. 1, where conductance is finite but has a value less than its maximum at $E_f$, *e.g.* <$14G_0$ for 7-tube bundles. Fig. 1 illustrates that the reduction of conductance G displays a universal behavior, independent of the details of the tubes orientation inside the bundle. Furthermore, by averaging results from more than 10 configurations, the randomly-orientated samples give almost the same conductance spectra as that of the tubes aligned at fixed angles (Fig.1). We note that an individual (*n,n*) tube with *n≠3q* (*q*=integer) has a different rotational symmetry around its axis from $C_3$ symmetry of the triangular lattice for the tubes inside a bundle.

When *n=3q*, individual (*n,n*) tubes contain the $C_3$ symmetry, which is compatible with the triangular bundle lattice. Such compatibility makes these bundles to behave very differently than the *n≠3q* bundles. To be specific, we show the calculated conductance spectra of (*3,3*) and (*6,6*) SWNT bundles in Fig. 2. Depending on the rotational angle, the conductance spectra along the tube axis can have complete gaps, partial gaps, or no gap at all at Fermi level. If there is no gap, the bundle's conductance scales with the number of tubes inside the bundle, *i.e.* $G = 14G_o$ for bundle with seven tubes, reaching the maximal possible conductance. However, for (*3,3*) and (*6,6*) bundles having random tube orientations, the conductance spectra are rather similar to those of the *n≠3q* bundles shown in Fig. 2 as dotted lines. These anomalous results are generally observed in the bundles preserving three-fold rotational symmetry[15, 16], such as (3,3), (6,6), (9,9) nanotubes and so on.

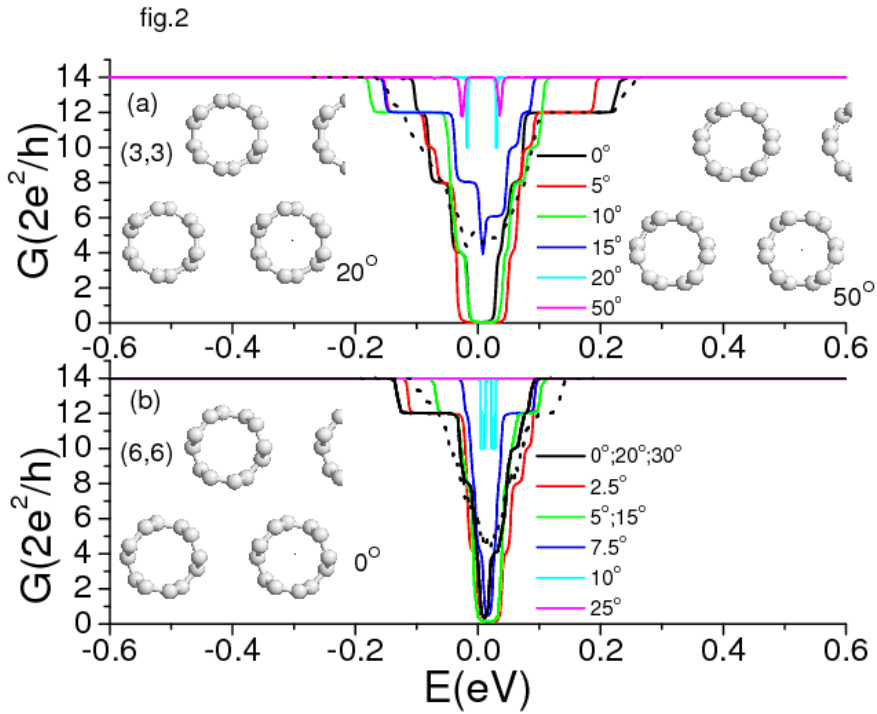

**Figure 2 (Color online) Conductance of bundles of with *n=3q*, (*3,3*) and (*6,6*) SWNT bundles. (a) 20 ° and 50 ° for (3,3) bundles and (b) 10 ° and 50 ° for (6,6) bundles are gapless at $E_f$. Other angles lead to either band gaps or pseudogaps at $E_f$. Dotted line: conductance spectra for averaged misaligned (*3,3*) and (*6,6*) bundle.**

More complex and diversified transport characteristics for the *n=3q* bundles can be well understood from symmetry arguments. In particular, we find the opening/closing of band gaps to originate from atomistic level coupling, by checking the detailed orientation configuration of the tubes in the bundle against the relevant electronic structures. There are two types of in-registry modes that allow maximum compatibility between the rotational symmetry of a tube around its axis and the global symmetry of the bundle's triangular lattice. If the symmetry of the single tube is compatible with the space group of the bundle lattice, the bundle can retain high symmetries, and thus the states crossing the Fermi level can transform according to different irreducible representations of the bundle space group. In the highest symmetry configurations, avoided crossings for π and π* bands can disappear at $E_f$, and the bundles attain maximal conductance. Other configurations with less compatibility have band gap opening due to tube-tube interactions and if the gap is big enough (depending on the coupling strength), the number of available states at $E_f$ would be suppressed. We have confirmed this symmetry picture by additional studies of a hypothetical square-lattice bundle. The commensurability condition also requires compatible in-registry rotational angles of the individual tubes for the possible conductance maximum for tubes with 4-fold symmetries, such as (4,4) tubes.

For any (*n,n*) bundle with *n=3q* in a triangular lattice arrangement, we find the first two highest commensurate tubular rotational angles to be $60°/n$ and $150°/n$ (or equivalently -$30°/n$). Therefore in Fig. 2 the $20°$ and $50°$ rotations for (*3,3*), the $10°$ and $25°$ rotations for (*6,6*) bundles have no reduction of conductance at $E_f$ and their conductance reach the maximum $14G_o$ at $E_f$. We identify the aligned *n=3q* bundles (if *n* is an even number) can attain high symmetries such as P6/MMM(D6H-1) and P6/MCC(D6H-2); while for aligned odd *n* bundles, there can be P63/MMC(D6H-4) and P63/MCM(D6H-3) symmetries. For incommensurate out-of-registry angles, lower symmetries such as P6/M(C6H-1) for even *n* and P63/M(C6H-2) for odd *n* are found. Randomly oriented bundles for both even and odd *n* generally have the lowest P1(C1) symmetry only.

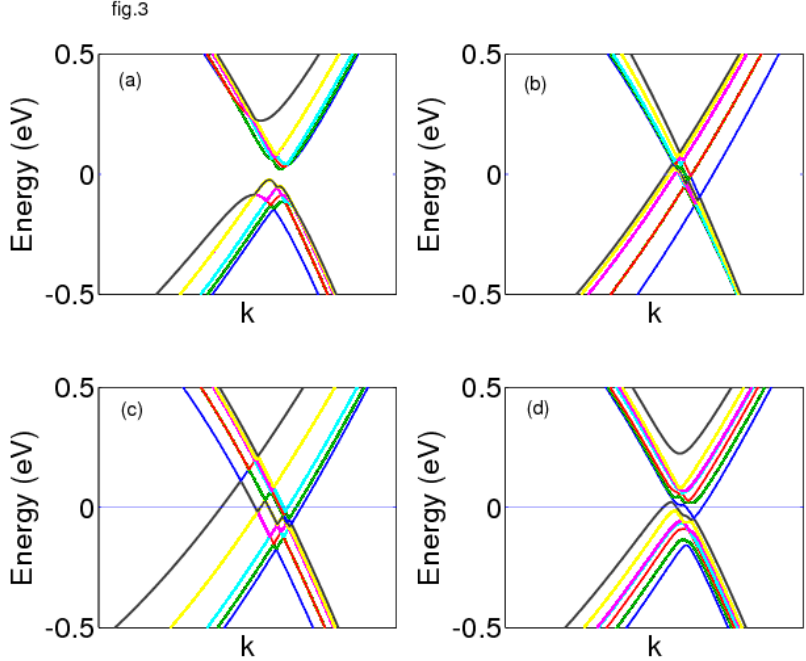

**Fig. 3 (Color online) Band structures for aligned (3,3) bundle with rotational angle 0, 20, 50° and randomly-orientated (a-d respectively). Complete and partial anti-crossings of seven linear dispersion bands for the seven armchair SWNTs in the bundle are shown.**

The conductance spectra are well correlated with the band structures. In a nanotube bundle, each tube should contribute one pair (non-degenerate) $\pi^*$ and $\pi$ bands as shown in Fig. 3 for bundle with seven tubes. For a *general* given incommensurate collective rotational angle, *e.g.* $0°$ *(3,3)* aligned bundle in Fig. 3a, all the constituent bands cannot cross the Fermi level. These avoided crossings give rise to a complete band gap, and the bundle has zero conductance. For the *specific* commensurate rotational angles like $20°$ and $50°$ (Fig.3b and 3c), the high symmetry allows the persistence of band crossings for all of the bands at $E_f$. The absence of band gap at $E_f$ then gives exactly 2N (N=7 is the tube number) channels of maximal conductance. For randomly orientated bundles (Fig. 3d), there is a coexistence of band crossings and repulsions, showing semimetal-like DOS and 1/3 maximal conductance. When $n \neq 3q$, the bundles' band structures are not sensitive to the rotational angles for all bundles, due to the general loss of commensurate symmetries in these tube bundles. This is the reason why similar DOS pseudogaps could be obtained for both aligned and misaligned (*10,10*) SWNTs[3, 17].

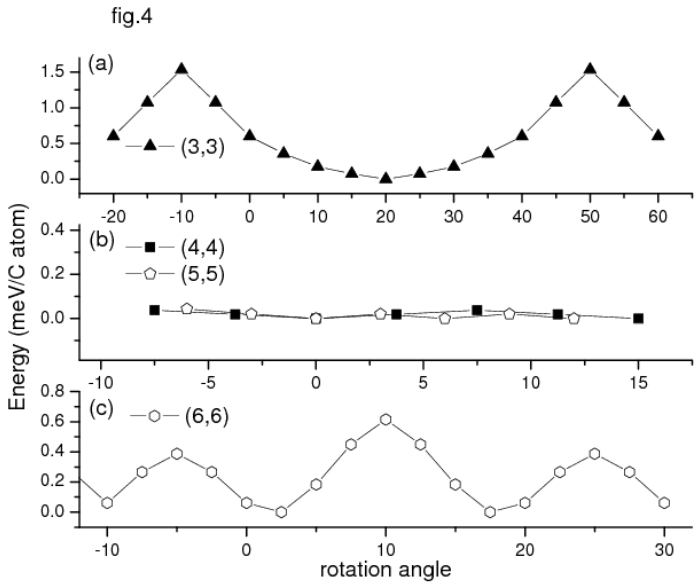

**Figure 4. Total energies per carbon atom for bundles are illustrated for (3,3) to (6,6) tubes, relative to the lowest energy orientation (taken as zero energy reference).**

Total energies calculated within DFT[18] for (*3,3*) and (*6,6*) bundles exhibit strong dependence on the rotational angles of tubes, while those of (4,4) and (5,5) are nearly independent of orientation, as shown in Fig 4. For (*3,3*) bundles, the two "gapless" angles have maximal ($50^o$ or equivalently $-10^o$) and minimal ($20^o$) energies; while for (*6,6*) bundles, one has the maximal ($10^o$) energy and the other ($25^o$ or $-5^o$) has the second highest energy. We can trace this angle-dependent energetics to the delicate stacking of neighboring tube walls[15]. The total energies depend on the closest wall areas of two neighboring tubes. We found that the high energy configurations generally correspond to an AA-like stacking of hexagons on tube walls facing each other. The results shown in Fig. 4 also imply that if $n \neq 3q$, the dependence of total energy on orientation is absent, in full agreement with the universal conductance suppression. While for the case of $n=3q$, to reach the conductance maximum an alignment in some specific higher energy orientation can be necessary.

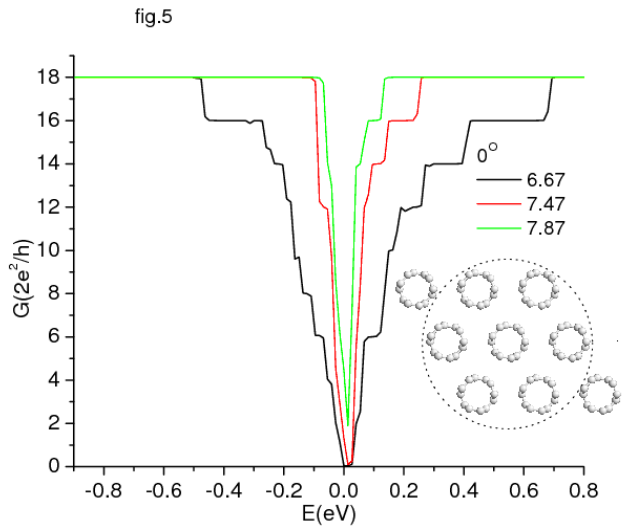

**Figure 5 (Color online) Conductance of relaxed bundles of nine (3,3) tubes with varying tube lattice constants (center-to-center). The band gap is inversely proportional to the intertube spacing (rotational angle 0º is shown).**

We also considered the effect of the bundle size and spacing between neighboring tubes (the latter is done in view of the fact that DFT may not be accurate in predicting the inter-tube distance of the bundle). The inset to Fig. 5 shows a nine-tube (*3,3*) aligned (0º) bundle whose maximum conductance is $18G_0$ and it has qualitatively same conductance spectra as the seven-tube bundle of Fig. 2a. This suggests that the number of tubes in the bundle is not a determining factor compared with the symmetry. But the distance between tubes in a bundle strongly affects the coupling, which in turn controls the size of the gaps near $E_f$ and the conductance. As the tubes are separated farther away from each other, the conductance gap gradually closes up when the wall-to-wall distance is ~5 Å (not shown). This is expected because that essentially the inter-tube coupling diminishes, and results in the disappearance of the conductance gap. From Fig. 5, it is interesting to observe that conductance of a finite-sized bundle in the pseudogap region still undergoes quantized changes, showing narrow conductance plateaus of integer multiples of $G_0=2e^2/h$. In particular, our calculation has indicated that each of these conductance plateaus is characterized by a van Hove peak in DOS at energies corresponding to the conductance plateau edge. On the other hand, when infinite number of bundles forms a periodic crystal, DOS in the pseudogap region becomes smooth[3] without the van Hove singularities of the one dimensional small bundle.

In conclusion for useful applications of SWNT bundles in interconnect technology, the conductance should be large and predictable (i.e. not dependent on details). We have identified the geometric conditions by which conductance maximum can be obtained at the Fermi energy of the bundle. If the tubes are not of the *n=3q* type, the conductance are rather determined. From both energetic and transport considerations, we obtain one third conductance of N single tubes. In contrast when *n=3q* these conditions appear to be very delicate by depending on rotational angles of individual tubes and sometimes, the most stable total energy configuration can correspond to nonoptimal conductance. Separating individual tubes laterally could eliminate this deterioration due to the bundling effects. On the other hand, one desires a small inter-tube distance for packing more tubes in a given area to enhance current density. These conflicting requirements are also crucial in the case if we pack semiconducting and metallic tubes into bundles containing a distribution of diameters or chiralities[19].

W.R. is under financial support of HKUST through RPC06/07.SC21. T.H.C and T.C.L. acknowledge NCTS, NSC of Taiwan under Grant No.NSC-96-2112-M194-012-MY3 and National Center for High Performance Computing. We also acknowledge financial support of HKSAR RGC grants (CA04/05.SC02) (P.S.), (HKU 7048/06P) (J.W.), and NSERC of Canada (H.G.). The computation resources were supported by the Shun Hing Education and Charity Fund. We thank Professor Steven G. Louie for discussions.